# Introduction to Relativistic Collisions

Frank W. K. Firk
*The Henry Koerner Center for Emeritus Faculty, Yale University*
*New Haven CT 06520*

*Abstract*: This introduction is intended for students taking a first course in Nuclear and Particle Physics. Electron-positron annihilation-in-flight is discussed as a topical example of an exact relativistic analysis.

## 1. 4-momentum and the energy-momentum invariant

In Classical Mechanics, the concept of momentum is important because of its rôle as an invariant in an isolated system. We therefore introduce the concept of 4-momentum in Relativistic Mechanics in order to find possible Lorentz invariants involving this new quantity. The contravariant 4-momentum is defined as:

$$P^\mu = mV^\mu \qquad (1)$$

where m is the mass of the particle (it is a Lorentz scalar — the mass measured in the rest frame of the particle), and $V^\mu[\gamma c, \gamma \mathbf{v}_N]$ is the 4-velocity. Here, $\gamma$ is the relativistic factor and $\mathbf{v}_N$ is the Newtonian 3-velocity,

Now, $V^\mu V_\mu = c^2$, where c is the invariant speed of light, therefore

$$P^\mu P_\mu = (mc)^2. \qquad (2)$$

The 4-momentum can be written,

$$P^\mu = [m\gamma c, m\gamma \mathbf{v}_N] \qquad (3)$$

therefore,

$$P^\mu P_\mu = (m\gamma c)^2 - (m\gamma \mathbf{v}_N)^2.$$

Writing

M = γm, the relativistic mass, we obtain

$$P^\mu P_\mu = (Mc)^2 - (M\mathbf{v}_N)^2 = (mc)^2. \quad (4)$$

Multiplying throughout by $c^2$ gives

$$M^2c^4 - M^2v_N^2c^2 = m^2c^4. \quad (5)$$

The quantity $Mc^2$ has dimensions of energy; we therefore write

$$E = Mc^2, \quad (6)$$

the total energy of a freely moving particle.

*This leads to the fundamental invariant of dynamics*

$$c^2 P^\mu P_\mu = E^2 - (\mathbf{p}c)^2 = E^{o2} \quad (7)$$

where

$E^o = mc^2$ is the rest energy of the particle, and **p** is its *relativistic 3-momentum*.

The total energy can be written:

$$E = \gamma E^o = E^o + T, \quad (8)$$

where

$$T = E^o(\gamma - 1), \quad (9)$$

the *relativistic kinetic energy*.

The magnitude of the 4-momentum is a Lorentz invariant

$$|P^\mu| = mc. \tag{10}$$

The 4-momentum transforms as follows:

$$P'^\mu = \mathbf{L}P^\mu, \text{ where } \mathbf{L} \text{ is the Lorentz operator.} \tag{11}$$

## 2. The relativistic Doppler shift

For relative motion along the x-axis, the equation $P'^\mu = \mathbf{L}P^\mu$ is equivalent to the equations (here, $\beta = v/c$)

$$E' = \gamma E - \beta\gamma cp^x \tag{12}$$

and,

$$cp'^x = -\beta\gamma E + \gamma cp^x. \tag{13}$$

Using the Planck-Einstein equations $E = h\nu$ and $E = p^x c$ for photons, the energy equation becomes

$$\nu' = \gamma\nu - \beta\gamma\nu$$

$$= \gamma\nu(1 - \beta)$$

$$= \nu(1 - \beta)/(1 - \beta^2)^{1/2}$$

$$= \nu\{(1 - \beta)/(1 + \beta)\}^{1/2}. \tag{14}$$

This is the relativistic Doppler shift for the frequency $\nu'$, measured in an inertial frame (primed) in terms of the frequency $\nu$ measured in another inertial frame (unprimed).

## 3. Relativistic collisions and the conservation of 4-momentum

Consider the interaction between two particles, 1 and 2, to form two particles, 3 and 4. (3 and 4 are not necessarily the same as 1 and 2). The contravariant 4-momenta are $P_i^\mu$ :

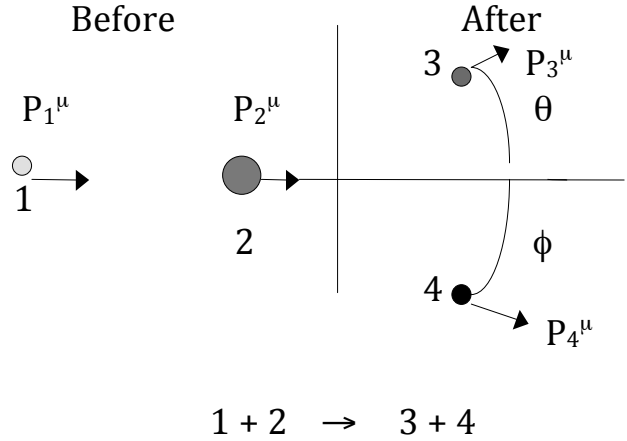

$$1 + 2 \rightarrow 3 + 4$$

All experiments are consistent with the fact that the 4-momentum of the system is conserved. We have, for the contravariant 4-momentum vectors of the interacting particles.

$$P_1^\mu + P_2^\mu = P_3^\mu + P_4^\mu \qquad (15)$$

$$\underbrace{\phantom{P_1^\mu + P_2^\mu}}_{\uparrow\text{ initial "free" state}} \quad \underbrace{\phantom{P_3^\mu + P_4^\mu}}_{\uparrow\text{ final "free" state}}$$

and a similar equation for the covariant 4-momentum vectors,

$$P_{1\mu} + P_{2\mu} = P_{3\mu} + P_{4\mu} . \qquad (16)$$

If we are interested in the change $P_1^\mu \rightarrow P_3^\mu$, then we require

$$P_1^\mu - P_3^\mu = P_4^\mu - P_2^\mu \qquad (17)$$

and

$$P_{1_\mu} - P_{3_\mu} = P_{4_\mu} - P_{2_\mu}. \tag{18}$$

Forming the invariant scalar products, and using $P_{i_\mu}P_i^\mu = (E_i^0/c)^2$, we obtain

$$(E_1^0/c)^2 - 2(E_1E_3/c^2 - \mathbf{p}_1\cdot\mathbf{p}_3) + (E_3^0/c)^2$$

$$= (E_4^0/c)^2 - 2(E_2E_4/c^2 - \mathbf{p}_2\cdot\mathbf{p}_4) + (E_2^0/c)^2. \tag{19}$$

Introducing the scattering angles, $\theta$ and $\phi$, this equation becomes

$$E_1^{0\,2} - 2(E_1E_3 - c^2p_1p_3\cos\theta) + E_3^{0\,2}$$

$$= E_2^{0\,2} - 2(E_2E_4 - c^2p_2p_4\cos\phi) + E_4^{0\,2}.$$

If we choose a reference frame in which particle 2 is at rest (the LAB frame), then $\mathbf{p}_2 = 0$ and $E_2 = E_2^0$, so that

$$E_1^{0\,2} - 2(E_1E_3 - c^2p_1p_3\cos\theta) + E_3^{0\,2} = E_2^{0\,2} - 2E_2^0E_4 + E_4^{0\,2} \tag{20}$$

The total energy of the system is conserved, therefore

$$E_1 + E_2 = E_3 + E_4 = E_1 + E_2^0 \tag{21}$$

or

$$E_4 = E_1 + E_2^0 - E_3$$

Eliminating $E_4$ from the above "scalar product" equation gives

$$E_1^{0\,2} - 2(E_1E_3 - c^2p_1p_3\cos\theta) + E_3^{0\,2} = E_4^{0\,2} - E_2^{0\,2} - 2E_2^0(E_1 - E_3). \tag{22}$$

This is the basic equation for all interactions in which two relativistic entities in the initial state interact to give two relativistic entities in the final state. It applies equally well to interactions that involve massive and massless entities.

## 3.1 The Compton effect

The general method discussed in the previous section can be used to provide an exact analysis of Compton's famous experiment in which the scattering of a photon by a stationary, free electron was studied. In this example, we have

$E_1 = E_{ph}$ (the incident photon energy), $E_2 = E_e^0$ (the rest energy of the stationary electron, the "target"), $E_3 = E_{ph}'$ (the energy of the scattered photon), and $E_4 = E_e'$ (the energy of the recoilling electron). The "rest energy" of the photon is zero:

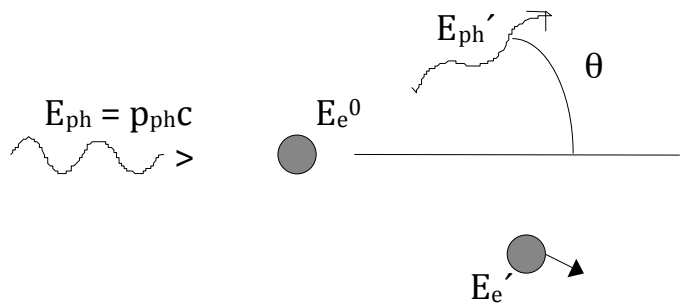

The general equation is now

$$0 - 2(E_{ph}E_{ph}' - E_{ph}E_{ph}'\cos\theta) = E_e^{0\,2} - 2E_e^0(E_{ph} + E_e^0 - E_{ph}') + E_e^{0\,2} \quad (23)$$

or

$$-2E_{ph}E_{ph}'(1 - \cos\theta) = -2E_e^0(E_{ph} - E_{ph}')$$

so that

$$E_{ph} - E_{ph}' = E_{ph}E_{ph}'(1 - \cos\theta)/E_e^0 . \qquad (24)$$

Compton measured the energy-loss of the photon on scattering and its $\cos\theta$-dependence.

## 4. Relativistic inelastic collisions

We shall consider an inelastic collision between a particle 1 and a particle 2 (initially at rest) to form a composite particle 3. In such a collision, the 4-momentum is conserved (as it is in an elastic collision) however, the kinetic energy is not conserved. Part of the kinetic energy of particle 1 is transformed into excitation energy of the composite particle 3. This excitation energy can take many forms — heat energy, rotational energy, and the excitation of quantum states at the microscopic level. The inelastic collision is as shown:

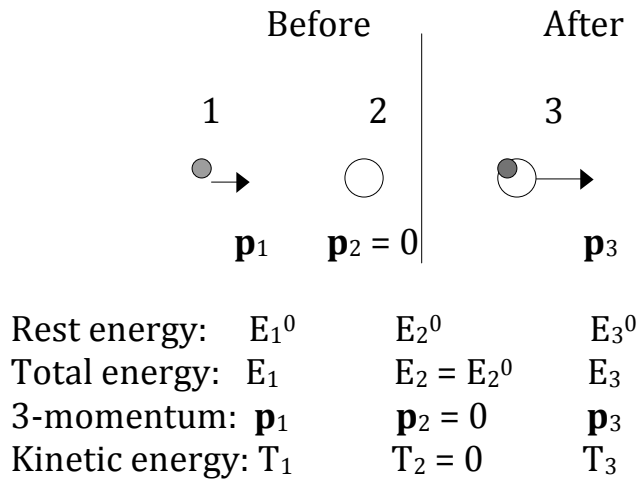

|  | Before | | After |
|---|---|---|---|
|  | 1 | 2 | 3 |
|  | $\mathbf{p}_1$ | $\mathbf{p}_2 = 0$ | $\mathbf{p}_3$ |
| Rest energy: | $E_1^0$ | $E_2^0$ | $E_3^0$ |
| Total energy: | $E_1$ | $E_2 = E_2^0$ | $E_3$ |
| 3-momentum: | $\mathbf{p}_1$ | $\mathbf{p}_2 = 0$ | $\mathbf{p}_3$ |
| Kinetic energy: | $T_1$ | $T_2 = 0$ | $T_3$ |

In this problem, we shall use the energy-momentum invariants associated with each particle, directly:

i) $E_1^2 - (\mathbf{p}_1 c)^2 = E_1^{0\ 2}$ (25)

ii) $E_2^2 \qquad\quad = E_2^{0\ 2}$ (26)

iii) $E_3^2 - (\mathbf{p}_3 c)^2 = E_3^{0\ 2}$. (27)

The total energy is conserved, therefore

$$E_1 + E_2 = E_3 = E_1 + E_2^0 \,. \tag{28}$$

Introducing the kinetic energies of the particles, we have

$$(T_1 + E_1^0) + E_2^0 = T_3 + E_3^0 \,. \tag{29}$$

The 3-momentum is conserved, therefore

$$\mathbf{p}_1 + 0 = \mathbf{p}_3 \,. \tag{30}$$

Using

$$E_3^{0\ 2} = E_3^2 - (\mathbf{p}_3 c)^2, \tag{31}$$

we obtain

$$\begin{aligned}
E_3^{0\ 2} &= (E_1 + E_2^0)^2 - (\mathbf{p}_3 c)^2 \\
&= E_1^2 + 2E_1 E_2^0 + E_2^{0\ 2} - (\mathbf{p}_1 c)^2 \\
&= E_1^{0\ 2} + 2E_1 E_2^0 + E_2^{0\ 2} \\
&= E_1^{0\ 2} + E_2^{0\ 2} + 2(T_1 + E_1^0)E_2^0
\end{aligned} \tag{32}$$

or

$$E_3^{0\,2} = (E_1^0 + E_2^0)^2 + 2T_1 E_2^0 \quad (E_3^0 > E_1^0 + E_2^0). \tag{33}$$

Using $T_1 = \gamma_1 E_1^0 - E_1^0$, where $\gamma_1 = (1 - \beta_1^2)^{-1/2}$ and $\beta_1 = v_1/c$, we have

$$E_3^{0\,2} = E_1^{0\,2} + E_2^{0\,2} + 2\gamma_1 E_1^0 E_2^0. \tag{34}$$

If two identical particles make a completely inelastic collision then

$$E_3^{0\,2} = 2(\gamma_1 + 1)E_1^{0\,2}. \tag{35}$$

## 5. The Mandelstam variables

In discussions of relativistic interactions it is often useful to introduce additional Lorentz invariants that are known as Mandelstam variables. They are, for the special case of two particles in the initial and final states $(1 + 2 \rightarrow 3 + 4)$:

$s = (P_1^\mu + P_2^\mu)[P_{1\mu} + P_{2\mu}]$, the total 4-momentum invariant

$$= ((E_1 + E_2)/c, (\mathbf{p}_1 + \mathbf{p}_2))[(E_1 + E_2)/c, -(\mathbf{p}_1 + \mathbf{p}_2)]$$

$$= (E_1 + E_2)^2/c^2 - (\mathbf{p}_1 + \mathbf{p}_2)^2 \tag{36}$$

→ Lorentz invariant,

$t = (P_1^\mu - P_3^\mu)[P_{1\mu} - P_{3\mu}]$, the 4-momentum transfer $(1 \rightarrow 3)$ invariant

$$= (E_1 - E_3)^2/c^2 - (\mathbf{p}_1 - \mathbf{p}_3)^2 \tag{37}$$

→ Lorentz invariant,

and

$$u = (P_1^\mu - P_4^\mu)[P_{1_\mu} - P_{4_\mu}], \text{ the 4-momentum transfer}$$

(1→4) invariant

$$= (E_1 - E_4)^2/c^2 - (\mathbf{p}_1 - \mathbf{p}_4)^2 \qquad (38)$$

→ Lorentz invariant.

Now,

$$sc^2 = E_1^2 + 2E_1E_2 + E_2^2 - (p_1^2 + 2\mathbf{p}_1\cdot\mathbf{p}_2 + p_2^2)c^2$$

$$= E_1^{0\ 2} + E_2^{0\ 2} + 2E_1E_2 - 2\mathbf{p}_1\cdot\mathbf{p}_2 c^2$$

$$= E_1^{0\ 2} + E_2^{0\ 2} + 2(E_1, \mathbf{p}_1 c)[E_2, -\mathbf{p}_2 c]. \qquad (39)$$

$$\underline{\phantom{xxxxxxx}}$$
$$\uparrow$$
Lorentz invariant

*The Mandelstam variable $sc^2$ has the same value in all inertial frames.* We therefore evaluate it in the LAB frame, defined by the vectors

$$[E_1^L, \mathbf{p}_1^L c] \text{ and } [E_2^L = E_2^0, -\mathbf{p}_2^L c = 0], \qquad (40)$$

so that

$$2(E_1^L E_2^L - \mathbf{p}_1^L\cdot\mathbf{p}_2^L c^2) = 2E_1^L E_2^0, \qquad (41)$$

and

$$sc^2 = E_1^{0\ 2} + E_2^{0\ 2} + 2E_1^L E_2^0. \qquad (42)$$

We can evaluate sc² in the center-of -mass (CM) frame, defined by the condition

$\mathbf{p}_1^{CM} + \mathbf{p}_2^{CM} = 0$ (the total 3-momentum is zero):

$$sc^2 = (E_1^{CM} + E_2^{CM})^2. \qquad (43)$$

This is the square of the total CM energy of the system.

## 5.1 The total CM energy and the production of new particles

The quantity $c\sqrt{s}$ is the energy available for the production of new particles, or for exciting the internal structure of particles. We can now obtain the relation between the total CM energy and the LAB energy of the incident particle (1) and the target (2), as follows:

$$sc^2 = E_1^{0\ 2} + E_2^{0\ 2} + 2E_1^L E_2^0 = (E_1^{CM} + E_2^{CM})^2 = W^2, \text{ say.} \qquad (6.44)$$

Here, we have evaluated the left-hand side in the LAB frame, and the right-hand side in the CM frame!

At very high energies, $c\sqrt{s} \gg E_1^0$ and $E_2^0$, the rest energies of the particles in the initial state, in which case,

$$W^2 = sc^2 \approx 2E_2^L E_2^0. \qquad (45)$$

The total CM energy, W, available for the production of new particles therefore depends on the square root of the incident laboratory energy. This result led to the development of colliding, or intersecting, beams of particles (such as protons and anti-protons) in order to produce sufficient energy to generate particles with rest masses typically 100 times the rest mass of the proton (~ $10^9$ eV).

## 6. Positron-electron annihilation-in-flight

A discussion of the annihilation-in-flight of a relativistic positron and a stationary electron provides a topical example of the use of relativistic conservation laws. This process, in which two photons are spontaneously generated, has been used as a source of nearly monoenergetic high-energy photons for the study of nuclear photo-disintegration since 1960. The general result for a 1 + 2 → 3 + 4 interaction, given in section **3**, provides the basis for an exact calculation of this process; we have

$E_1 = E_{pos}$ (the incident positron energy), $E_2 = E_e^0$ (the rest energy of the stationary electron), $E_3 = E_{ph1}$ (the energy of the forward-going photon), and $E_4 = E_{ph2}$ (the energy of the backward-going photon). The rest energies of the positron and the electron are equal. The general equation now reads

$$E_e^{02} - 2\{E_{pos}E_{ph1} - cp_{pos}E_{ph1}(\cos\theta)\} + 0 = 0 - E_e^{02} - 2E_e^0(E_{pos} - E_{ph1}) \quad (46)$$

therefore

$$E_{ph1}\{E_{pos} + E_e^0 - [E_{pos}^2 - E_e^{02}]^{1/2} \cos\theta\} = (E_{pos} + E_e^0)E_e^0,$$

giving

$$E_{ph1} = E_e^0/(1 - k\cos\theta) \qquad (47)$$

where

$$k = [(E_{pos} - E_e^0)/(E_{pos} + E_e^0)]^{1/2}.$$

The maximum energy of the photon, $E_{ph1}^{max}$ occurs when $\theta = 0$, corresponding to motion in the forward direction; its energy is

$$E_{ph1}^{max} = E_{oe}/(1 - k). \qquad (48)$$

If, for example, the incident total positron energy is 30 MeV, and $E_e^0 = 0.511$ MeV then

$$E_{ph1}^{max} = 0.511/[1 - (29.489/30.511)^{1/2}] \text{ MeV}$$

$$= 30.25 \text{ MeV}.$$

The forward-going photon has energy equal to the kinetic energy of the incident positron ($T_1 = 30 - 0.511$ MeV) plus about three-quarters of the total rest energy of the positron-electron pair ($2E_e^0 = 1.02$ MeV). Using the conservation of the total energy of the system, we see that the energy of the backward-going photon is about 0.25 MeV.

The method of positron-electron annihilation-in-flight provides one of the very few ways of generating nearly monoenergetic photons at high energies. The method was first proposed by Tzara (1957).

# References

The following references place the subject of relativistic collisions in the general context of Special Relativity, and provide many specific examples of the application of the theory to experiments.